      [1999/12/01 v1.4c Il Nuovo Cimento]
\documentclass{cimento}

\usepackage{graphicx}  
\title{First results from the ALHAMBRA-Survey}
\author{A.~Fern\'andez-Soto\from{ins:1},
\\
on behalf of the ALHAMBRA-Survey Core Team\thanks{M. Moles (PI),
J.~A.~L. Aguerri, E.~J.~Alfaro, N. Ben\'{\i}tez, T. Broadhurst,
J.~Cabrera-Ca\~no, F.~J. Castander, J. Cepa, M. Cervi\~no,
D.~Crist\'obal-Hornillos, R.~M. Gonz\'alez Delgado, L. Infante ,
I. M\'arquez, V.~J. Mart\'{\i}nez, J. Masegosa, A. del Olmo, J. Perea,
F. Prada, J.~M. Quintana and S.~F. S\'anchez}}
\instlist{\inst{ins:1} Universitat de Val\`encia, Spain}

\PACSes{\PACSit{98.80}{(Cosmology)}}
\begin{document}

\maketitle

\begin{abstract}
We present the first results from the ALHAMBRA survey. ALHAMBRA will
cover a relatively wide area (4 square degrees) using a
purposely-designed set of 20 medium-band filters, down to an
homogeneous magnitude limit AB $\approx$ 25 in most of them, adding
also deep near-infrared imaging in $JHK_s$.  To this aim we are using
the Calar Alto 3.5m telescope. A small area of the ALHAMBRA survey has
already been observed through our complete filter set, and this allows
for the first time to check all the steps of the survey, including the
pipelines that have been designed for the project, the fulfilment of
the data quality expectations, the calibration procedures, and the
photometric redshift machinery for which ALHAMBRA has been
optimised. We present here the basic results regarding the properties
of the galaxy sample selected in a $15 \times 15$ arcmin$^2$ area of
the ALHAMBRA-8 field, which includes approximately 10000 galaxies with
precise photometric redshift measurements. In a first estimate,
approximately 500 of them must be galaxies with $z > 2$.
\end{abstract}

\section{Introduction}
The ALHAMBRA-Survey will produce accurate photometric redshifts for a
large number of objects, enough to track cosmic evolution, i.e., the
change with $z$ of the content and properties of the Universe, a kind
of {\em Cosmic Tomography}. ALHAMBRA is imaging 4 square degrees
(Table 1) with 20 contiguous equal-width medium-band filters
covering from 3500 \AA\ to 9700 \AA\ (see Figure 1), plus the standard
JHK$_s$ near-infrared bands. The optical photometric system has been
designed to maximise the number of objects with accurate
classification by Spectral Energy Distribution type and redshift and
to be sensitive to relatively faint emission features in the
spectrum~\cite{ref:ben}.  

\begin{table}[t]
\caption{The ALHAMBRA-Survey Fields}
\centering
\label{campos}
\begin{tabular}{|l|c|c|c|c|c|c|}
\hline
Field name  &  RA(J2000) & DEC(J2000)& 100 $\mu$m & E(B$-$V) & l & b \\
\hline
ALHAMBRA-1  & 00 29 46.0 & +05 25 28 & 0.83 & 0.017 &  113 & -57 \\
ALHAMBRA-2/DEEP2  & 02 28 32.0 & +00 47 00 & 1.48 & 0.031 & 166 & -53  \\
ALHAMBRA-3/SDSS  & 09 16 20 & +46 02 20 & 0.67 & 0.015 & 174 & +44 \\
ALHAMBRA-4/COSMOS  & 10 00 28.6 & +02 12 21 & 0.91 & 0.018 & 236 & +42 \\
ALHAMBRA-5/HDF-N   & 12 35 00.0 & +61 57 00 & 0.63 & 0.011 &  125 & +55 \\
ALHAMBRA-6/GROTH   & 14 16 38.0 & +52 25 05 & 0.49 & 0.007 &  95 & +60 \\
ALHAMBRA-7/ELAIS-N1 & 16 12 10.0 & +54 30 00 & 0.45 & 0.005 & 84 & +45 \\
ALHAMBRA-8/SDSS & 23 45 50.0 & +15 34 50 & 1.18 & 0.027 & 99 & -44 \\
\hline
\end{tabular}
\end{table}

\begin{figure}[b]
\centering
\includegraphics[angle=-90,width=11cm]{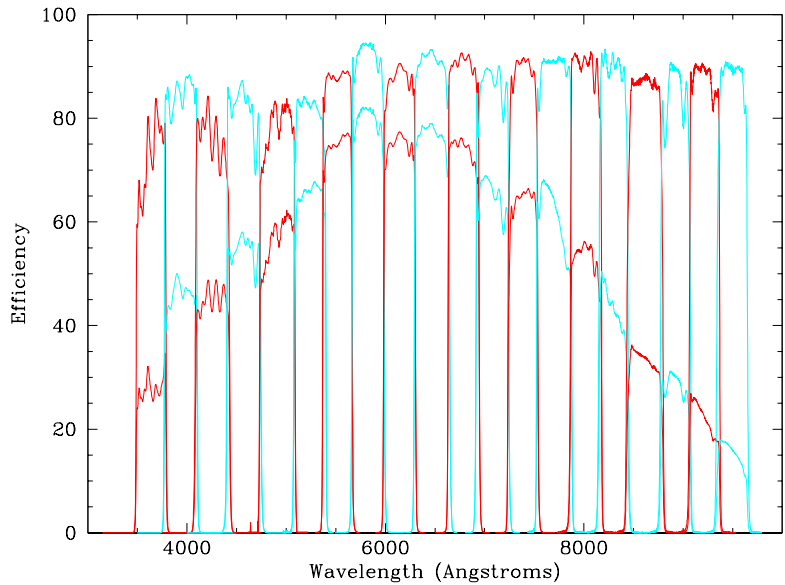}
\caption{Transmission curves measured for one of the optical filter
sets alone (higher curves), and combined with the joint effect of the
CCD detector, the atmosphere and the telescope (lower curves).}
\label{filtro}
\end{figure}

The observations are being carried out with the Calar Alto 3.5m
telescope using two wide field cameras: LAICA in the optical, and
OMEGA-2000 in the NIR. The magnitude limit, for a total of 100 ksec
integration time per pointing, is AB $\geq$ 25 mag (for an unresolved
object, S/N = 5) in the optical filters from the blue to 8300~\AA, and
ranges from AB = 24.7 to 23.4 for the redder ones. The limit in the
NIR, for a total of 15 ks exposure time per pointing, is K$_s$ = 20
mag, H = 21 mag, J = 22 mag. We expect to obtain accurate redshift
values $\Delta z/(1+z) \leq 0.03$ for about 6.6 $\times 10^5$ galaxies
with I$\leq 25$ (60\% completeness level) and $z_{med}$ = 0.74. In
Table 2 we compare the properties of the ALHAMBRA-Survey with other
similar endeavours in cosmic cartography.

\begin{table}[t]
\caption{Main characteristics of wide field ($\geq 0.5$ square
degrees) spectroscopic surveys} \centering
\label{Surveys}
\begin{tabular}{|l|c|c|c|c|}
\hline
Survey  &  Area (sqdeg) & Spectral range (\AA)& $z$ (median) & N$_{objects}
$ \\
\hline
CfA/SRSS & 18000 & 4300-6900 &  0.02 & 1.8 $\times$ 10$^4$\\
SDSS/DR6  &  6860 & 3800-9200 & 0.1 & 7.9 $\times$ 10$^5$ \\
LCRS      &  700  & 3350-6750 & 0.1 & 2.6 $\times$ 10$^4$\\
2dFGRS    &  2000 & 3700-8000 & 0.11 & 2.2 $\times$ 10$^5$ \\
VVDS    & 16 & 5500-9500 & 0.7 &  1.0 $\times$  10$^5$ \\
DEEP2  & 3.5  & 6500-9100 & 1.0 & 5.5 $\times$ 10$^3$  \\
\hline
ALHAMBRA-60 & 4 & 3500-9700 ($+JHK$) & 0.74 & 6.6 (3.0) $\times$ 10$^5$ \\
ALHAMBRA-90 & 4 & 3500-9700 ($+JHK$) & 0.63 & 3.5 (1.0) $\times$ 10$^5$ \\
\hline
\end{tabular}
\end{table}

This accuracy, together with the homogeneity of the selection
function, will allow for the study of the large scale structure
evolution with $z$, the galaxy luminosity function and its evolution,
the identification of clusters of galaxies, and many other studies,
without the need for any further follow-up. It will also provide
exciting targets for detailed studies with 10m class telescopes.
Given its area, spectral coverage and its depth, apart from those main
goals, ALHAMBRA will also produce valuable data for galactic studies.

\section{Redshift expectations}

We have calculated the number of objects that the survey will detect
to a fixed accuracy in the measured $z$-value, for the adopted
configuration and strategy. A more extense description of the
survey and initial results, is given in~\cite{ref:mol}. 

\begin{figure}[b]
\centering
\includegraphics[angle=0,width=9cm]{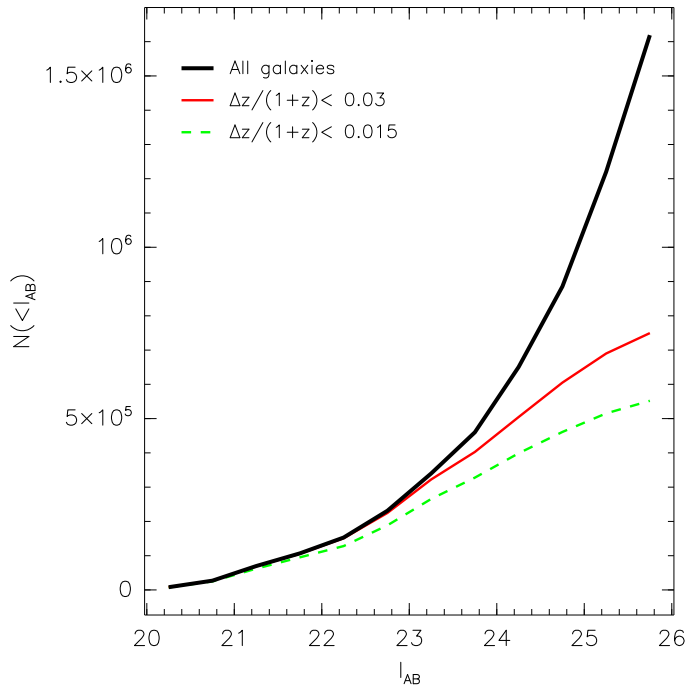}
\caption{The total number of galaxies detected in the
ALHAMBRA-Survey expected from our simulations (thick line). The thin
continuous (dashed) line gives the total number of galaxies with
$\Delta z/(1+z) \leq 0.03$ (0.015)} \label{Nmag}
\end{figure}

Amongst the different ways to present the figure of merit of a planned
survey, we show in Figure~\ref{Nmag} the total number of objects
detected to a given redshift accuracy. It can be seen that the total
number of objects with $\Delta z/(1+z) \leq 0.03$ is over 7$\times
10^5$, and it is over 5$\times 10^5$ for the higher accuracy of
0.015. Our simulations show that the survey will be complete at the
90\% level with $\Delta z/(1+z) \leq 0.03$ (0.015) down to I$\approx$
23.5 (21.8), and to the 60\% level till I$\approx$ 25.2 (24.3). In the
redshift-complete samples the number of objects expected are
3.5$\times 10^5$ (1.0 $\times 10^5$) and 6.6$\times 10^5$ (3.5$\times
10^5$) respectively. All the calculations were done using the package
BPZ~\cite{ref:bpz}.  Complete details about the simulations and
the results are given in~\cite{ref:ben}.

\section{Present status}

At the time of this conference, some of the project milestones have
already been reached and met with success. In particular, in August
2006 the first pointing (within the field ALHAMBRA-8) has been
completed, having been observed to the expected depth through all 20+3
filters. The percentage of the total area to be surveyed that has been
covered to date is approximately 55\% in the near infrared and 40\% in
the visible range---it must be remarked that OMEGA2000, the Calar Alto
NIR imager, was available for our observations from the very
beginning, whereas the optical imager Laica was still under
commissioning. We expect to complete our observations in 2009.

\begin{figure}
\centering
\includegraphics[angle=0,width=13cm]{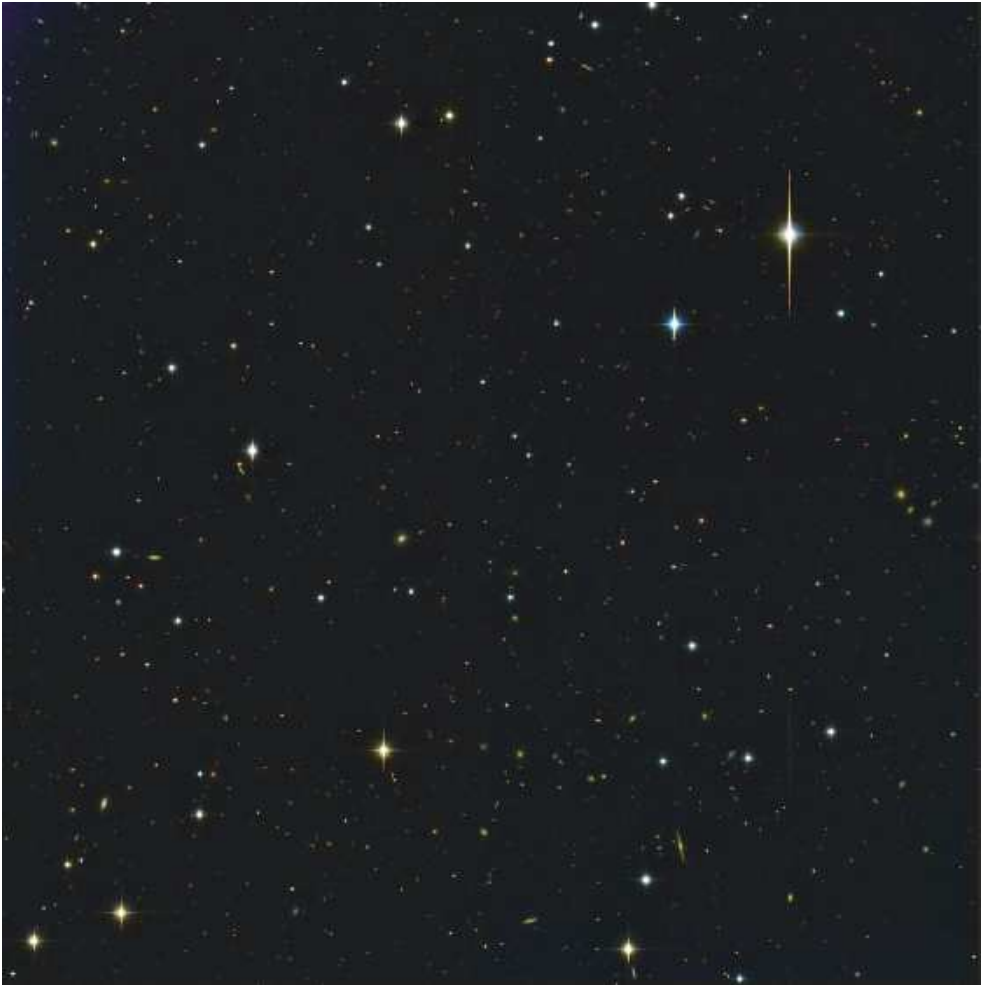}
\caption{The first complete pointing of the ALHAMBRA-8 field in a
square region of 15 arcminutes side. This color image has been created
making use of data from 14 out of the 23 filters. } \label{field}
\end{figure}

Our strategy has been from the beginning that of refraining from
starting the data analysis until the data treatment pipelines and
analysis tools are definitive. However, some basic analysis can be
performed with the first complete pointing, and this work is in fact
important for the definition and testing of the different tasks
involved in the survey. We show in Figure~\ref{field} a color image of
the first complete pointing in the ALHAMBRA-8 field. The region covers
an area of 15 arcminutes side and contains approximately 10000
objects to AB$\approx 25$.

Just to illustrate the depth that can be reached with the ALHAMBRA
images, we show in Figure 4 a small (1.3 $\times$ 1.8 arcminutes)
section of the ALHAMBRA-8 field. Here two bright red objects are
found. Each of the ten panels corresponds to one of the ALHAMBRA
filters running from F520 to F799, with central wavelengths that span
the range between 5200 and 7990 \AA. The reddest object (center left)
has a strong break between 6130 and 6440 \AA (AB(6130)$-$AB(6440) $>$
1.5 mag), and a second one between 7060 and 7370 \AA
(AB(7060)$-$AB(7370) $\sim$ 1 mag), which identify it as either a
moderate ($z \approx 1$) redshift galaxy with an old/reddened
population, or a high-redshift object at $z \approx 4$. A full,
calibrated ALHAMBRA ``spectrum'' can be seen in Figure 5, together
with the corresponding SDSS spectrum and photometry.

\begin{figure}
\centering
\includegraphics[angle=0,height=9cm]{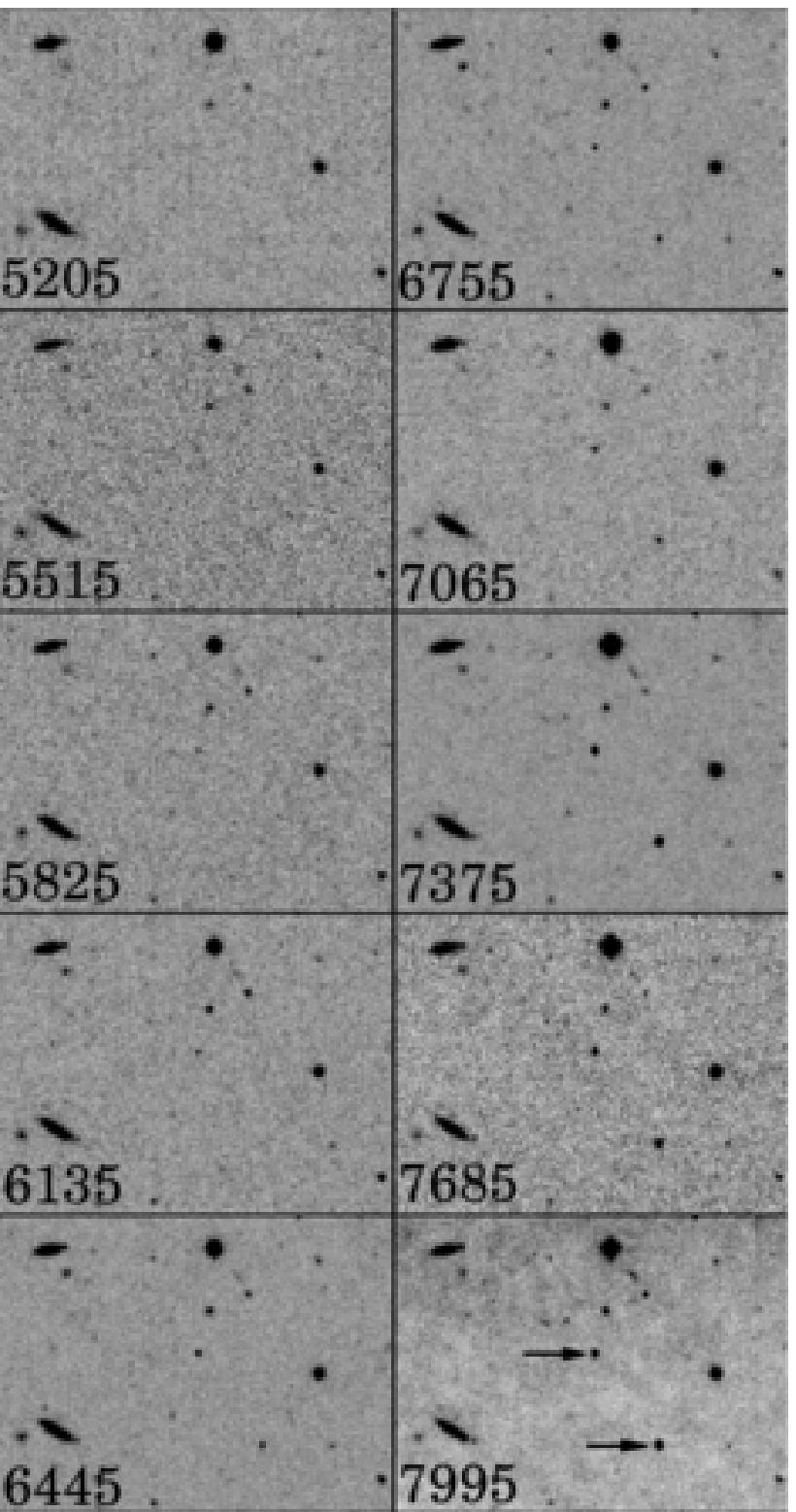}
\caption{A region of 1.3 $\times$ 1.8 arcminutes is shown in this
strip containing 10 images corresponding to 10 optical filters
spanning from 5200 to 7990 \AA. Two easily identified bright red
objects having $AB(7990)-AB(5200) > 3$ are marked in the bottom right
image, corresponding to $\lambda=7990$ \AA.}
\label{strip}
\end{figure}

\begin{figure}
\centering
\includegraphics[angle=-90,width=9cm]{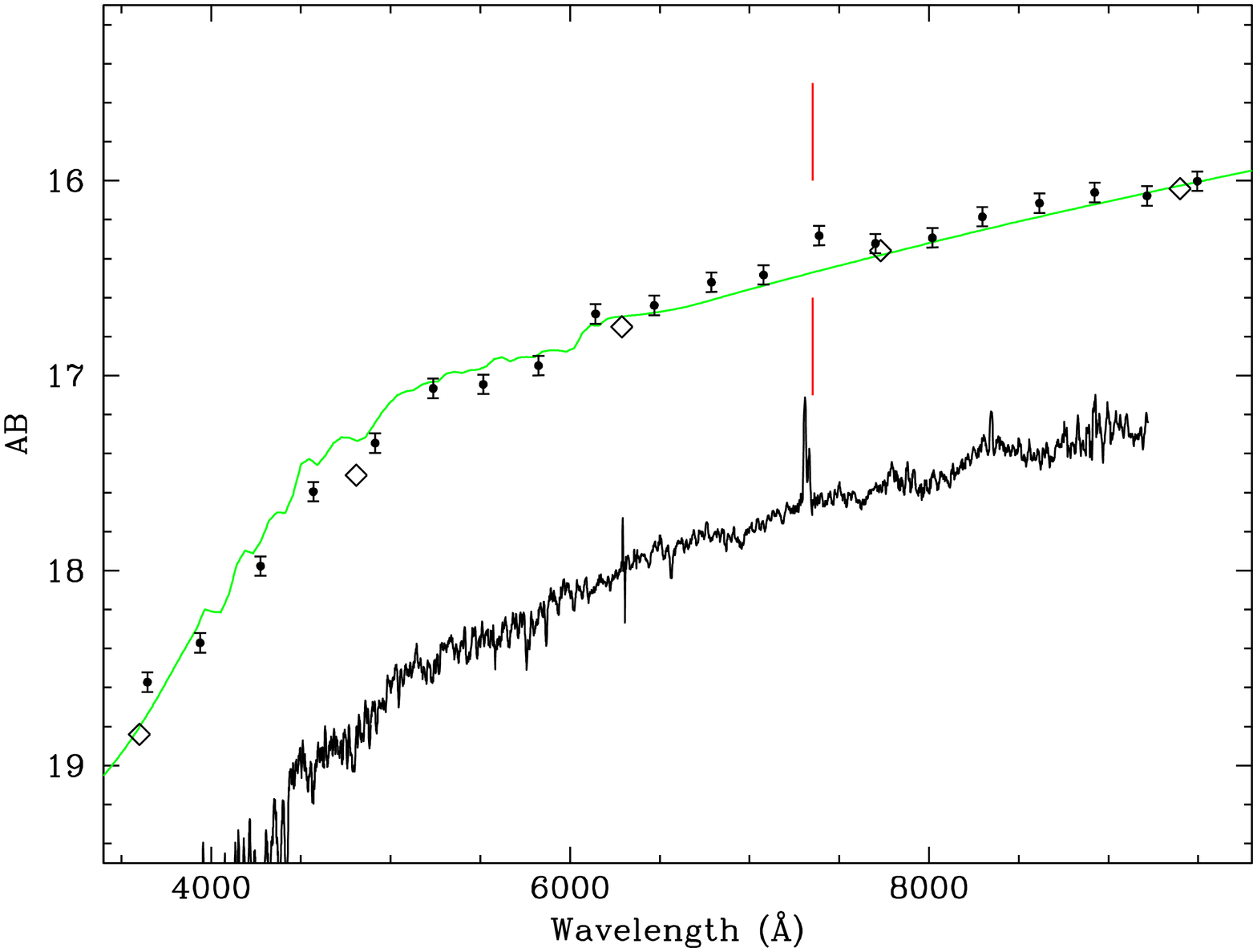}
\caption{SDSS spectrum (line) and photometry (diamonds), and ALHAMBRA 
photometry for one of the galaxies in our field with SDSS
spectroscopy. The vertical ticks mark the expected position of the
H$\alpha$ emission line at the redshift measured from the ALHAMBRA
data alone. The SDSS spectrum is $\approx1$ magnitude fainter because
it includes only the flux within the SDSS spectrograph fiber.}
\end{figure}

\acknowledgments
We acknowledge the decisive support given by the ALHAMBRA Extended
Team to the project (see http://alhambra.iaa.csic.es:8080/ for the
details regarding the project implementation and organisation). We
also wish to acknowledge the Calar Alto staff for their warm
assistance for a fruitful observation. The Ministerio de Educaci\'on y
Ciencia is acknowledged for its support through the grants
AYA2002-12685-E and AYA2004-20014-E, and project AYA2006-14056.

\end{document}